\documentclass[twocolumn,prb,showpacs,preprintnumbers,amsmath,amssymb]{revtex4-1} 

\usepackage{graphicx}
\usepackage{dcolumn}
\usepackage{bm}
\usepackage{verbatim}

\begin{document}


\title{Angular slippage from the crystallographic $c$-axis of the reversible magnetization vector in a tilted crystal of a highly anisotropic cuprate superconductor}

\author{J. Mosqueira$^1$}
 \email{j.mosqueira@usc.es}
\author{R.I. Rey$^1$}
\author{A. Wahl$^2$}
\author{F. Vidal$^1$}

\affiliation{$^1$LBTS, Facultade de F\'isica, Universidade de Santiago de Compostela, E-15782 Santiago de Compostela, Spain\\
$^2$Laboratoire CRISMAT, ENSICAEN, UMR 6508 CNRS, Caen, France}


\begin{abstract}
The magnetization \textit{vector} $\vec M$ was measured in the reversible region of the mixed state of a high quality Tl$_2$Ba$_2$Ca$_2$Cu$_3$O$_{10}$ single crystal as a function of temperature and for different magnetic field amplitudes and orientations.
These measurements allowed to study the $\vec M$ components perpendicular and parallel to the CuO$_2$ layers ($M_\perp$ and $M_\parallel$, respectively) under arbitrary values for the corresponding components of the applied magnetic field, $H_\perp$ and $H_\parallel$. 
For temperatures close to $T_c$ (in the critical fluctuation region) we observed $M_\perp(H_\perp,H_\parallel)\approx M_\perp(H_\perp,0)$ and $M_\parallel(H_\perp,H_\parallel)\approx0$, as expected for an extremely anisotropic material. However, deviations from this behavior are observed at lower temperatures in the London region. In particular, the $M_{\perp}$ amplitude under a constant $H_{\perp}$ decreases on increasing $H_{\parallel}$. 
In turn, in spite of the experimental uncertainties affecting $M_\parallel$ (mainly associated with the rotating sample holder), its amplitude under a constant $H_{\parallel}$ becomes observable on increasing $H_{\perp}$. 
Both effects lead to an angular slippage of the magnetization vector $\vec M$ from the $c$ crystallographic axis when the applied magnetic field is tilted from that axis.
The $M_{\perp}$ behavior is phenomenologically explained in the framework of Lawrence-Doniach approaches for single layered superconductors by just assuming a slight angular dependence of the so-called \textit{vortex structure constant}. However, the observed $M_\parallel$ is orders of magnitude larger than expected, a fact which is related to the multilayered nature of the compound studied. Our present results also directly affect the interpretation of recent measurements of the magnetic torque in other extremely anisotropic high-$T_c$ cuprates.

\end{abstract}

\maketitle

\section{introduction}

The magnetization in the reversible region of high-$T_c$ cuprate superconductors (HTSC) is a very useful tool to probe the nature of the superconductivity in these materials.\cite{tinkham} 
Its phenomenology in the presence of magnetic fields applied perpendicular to the CuO$_2$ layers has been extensively studied,\cite{reviews,kogan81,koganclem81,reviewfeinberg,farrell88,bruckental06,obaidatPRB97,obaidat99,bruckental08,bruckental06,tuominen2,zech,liu91,liu92,obaidatPC97,hasan,zhukov, bugoslavskyPRB97,tuominen1,pugnat} including the behavior in the presence of tilted magnetic fields.
In this case, due to the strong anisotropy of these materials, the reversible magnetization vector $\vec M$  lies almost perpendicular to the CuO$_2$ layers in a wide angular region, and a component transverse to the applied magnetic field ($M_T$) appears in addition to the \textit{longitudinal} component ($M_L$).\cite{kogan81,koganclem81,reviewfeinberg} Such a transverse component has been studied through measurements of the magnetic torque,\cite{farrell88} but there are very few works where the two components of the magnetization \textit{vector} are simultaneously studied. 
Some of them focus on irreversible properties of the tilted vortex lattice, as its dynamics upon rotation,\cite{bruckental06,obaidatPRB97,obaidat99,bruckental08} its relaxation,\cite{bruckental06,tuominen2,zech}, or the pinning forces acting on it.\cite{liu91,liu92,obaidatPC97,hasan} Other works address effects associated with the layered superconducting structure appearing when the applied magnetic field is  slightly tilted from the CuO$_2$ layers, as the vortex lock-in to these layers.\cite{zhukov, bugoslavskyPRB97} However, only in Refs.~\onlinecite{bruckental06}, \onlinecite{tuominen1} and \onlinecite{pugnat} simultaneous $M_T$ and $M_L$ measurements in the reversible mixed state are presented, and they correspond to samples with a relatively moderate anisotropy. 

In this work we present detailed measurements of the temperature and magnetic field dependences of $\vec M$ in a high quality highly anisotropic Tl$_2$Ba$_2$Ca$_2$Cu$_3$O$_{10}$ (Tl-2223) single crystal. These measurements were performed in the reversible region of the $H$-$T$ phase diagram by using different crystal orientations with respect to the applied magnetic field. This allowed us to study the components of the magnetization in the directions perpendicular ($M_\perp$) and parallel ($M_\parallel$) to the \textit{ab} layers, under arbitrary values for the corresponding components of the magnetic field ($H_\perp$ and $H_\parallel$, respectively). 
In highly anisotropic superconductors, it is expected that 
\begin{equation}
M_\perp(\theta,H)\approx M_\perp(0,H_\perp),
\label{eq1}
\end{equation}
and 
\begin{equation}
|M_\parallel|\ll |M_\perp|,
\label{eq2}
\end{equation}
where $\theta$ is the angle between the crystal $c$ axis and the applied magnetic field.\cite{reviewfeinberg} These equations imply that $\vec M$ is perpendicular to the CuO$_2$ layers, and largely determined by the $\vec H$ component in that direction. This was recently confirmed near $T_c$ ($T\stackrel{>}{_\sim}0.85T_c$) in the same highly anisotropic HTSC.\cite{angular} However, here we show that farther below $T_c$, where the experimental uncertainty associated with the background contribution is negligible, Eqs.~(\ref{eq1}) and (\ref{eq2}) are no longer valid: $|M_\perp|$ for a constant $H_\perp$ is progressively reduced with the application of a parallel magnetic field. In turn, $M_\parallel$ presents finite values (well above the experimental uncertainties) when $H_\perp>0$. 
We show that the effect on $M_\perp$ may be explained at a phenomenological level in the framework of the well known approach by Bulaevskii, Ledvig and Kogan for highly anisotropic layered superconductors which includes the effect of thermal fluctuations of two-dimensional vortex positions.\cite{BLK} For that, it is just assumed that the so-called \textit{vortex-structure constant} presents a slight dependence of the magnetic field orientation, which may be related to a change in the effective size of the two-dimensional vortex cores with tilting.\cite{feinberg, bulaevskii92} However, the effect on $M_\parallel$ is orders of magnitude larger than the one predicted by theoretical approaches for single layered HTSC, which could be attributed to the multilayered nature of Tl-2223. 

An interesting consequence of our present results concerns the interpretation of measurements of the magnetic torque in terms of $M_\perp$: By assuming that Eqs.~(\ref{eq1}) and (\ref{eq2}) are applicable to highly anisotropic HTSC, the magnetic torque per unit volume in these materials may be approximated as 
\begin{equation}
\tau=\mu_0HM_\perp\sin\theta.
\label{torqueintro}
\end{equation}
This expression has been used in a number of papers to determine $M_\perp(T,H_\perp)$ directly from $\tau(T,H)$.\cite{martinez,drost,bergemann,naughton,li05,li10} In some of these works the resulting temperature and magnetic field dependences of $M_\perp$ were claimed to be beyond conventional Ginzburg-Landau (GL) descriptions for the superconducting transition in these materials.\cite{bergemann,naughton,li05,li10} In particular, in Refs.~\onlinecite{li05,li10} it is claimed that such a behavior supports the so-called \textit{vortex scenario} for the loss of long-range phase coherence at $T_c$.\cite{emery} However, these conclusions depend on the applicability of Eqs.~(\ref{eq1}) and (\ref{eq2}) to highly anisotropic HTSC, which is questioned by our present results. 

The paper is organized as follows: In Sec.~II we present experimental details about the crystal, the set-up of the rotating sample holder, the procedure to determine the background signal, and the results of the $M_L(T)_{\theta,H}$ and $M_T(T)_{\theta,H}$ measurements. In Sec.~III.A the experimental data are analyzed in the critical region near $T_c$, while in Sec.~III.B the analysis is extended to the London region of the reversible mixed state, where deviations from the behavior expected for highly anisotropic superconductors are clearly observed. In Sec.~III.C it is presented a review of data in the literature which could confirm our present results in other highly anisotropic HTSC. In Sec.~III.D the implications on the interpretation of recent high-field torque measurements are discussed. Finally, the general conclusions are presented in Sec.~IV.

%
%

\begin{figure}[b]
\includegraphics[scale=.5]{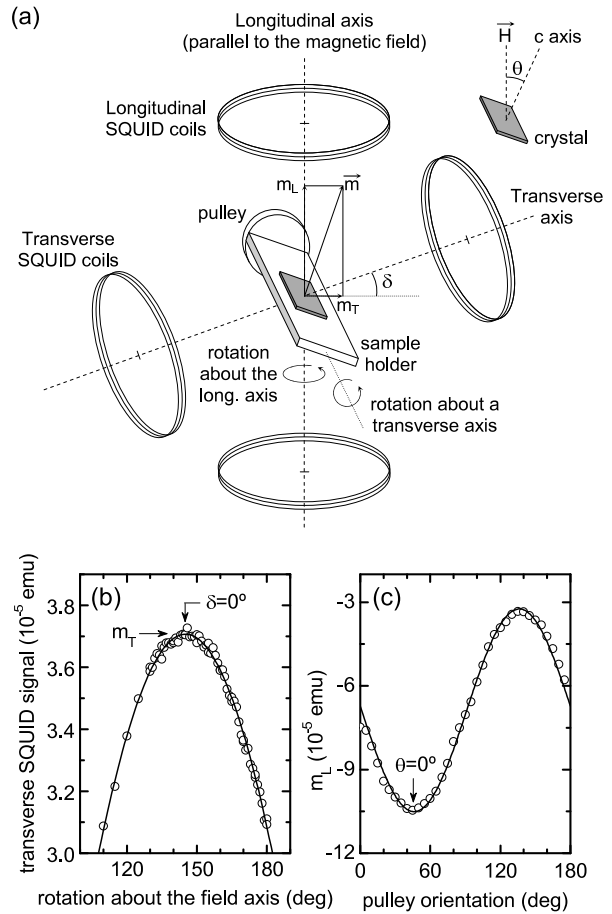}
\caption{(a) Schematic diagram of the rotating sample holder indicating the two rotation axes relative to the longitudinal and transverse SQUID coils. (b) Example of the procedure used to align $\vec m_T$ with the transverse SQUID coils by rotation about the longitudinal axis until the transverse SQUID signal was maximum. The line is a fit of ${\rm cte}+{\rm cte'}\cos\delta$, where $\delta$ is the misalignment angle. (c) Example of the $m_L$ dependence on the pulley orientation in the Meissner region. These measurements are used to determine the correspondence between the pulley orientation and $\theta$ (the angle between the crystal $c$ axis and the magnetic field). As shown in the main text, the minimum corresponds to $\theta=0^\circ$. The line is a fit of Eq.~(\ref{mltheta}) to the data.}
\label{esquema}
\end{figure}

\section{experimental details and results}

The Tl-2223 sample used in this work is a plate-like single crystal (1.1$\times$0.75$\times$0.226 mm$^3$) with the $c$ crystallographic axis perpendicular to the largest face. Details of its growth procedure may be seen in Ref.~\onlinecite{maignan}. Let us just mention that this crystal was already used in the magnetization measurements presented in Refs.~\onlinecite{angular} and \onlinecite{Tl2223}, and it has a sharp low-field diamagnetic transition ($T_c=122\pm1$~K), a large Meissner fraction ($\sim$80\%), and an excellent crystallinity (its mosaic spread is as low as $\sim0.1^\circ$).

The measurements were performed with a Quantum Design SQUID magnetometer, equipped with independent detectors for the components of the magnetic moment in the direction of the applied magnetic field (hereafter \textit{longitudinal}) and in a direction \textit{transverse} to it ($\vec m_L$ and $\vec m_T$ respectively). The sample was glued with a minute amount of GE varnish to a sample holder (also from Quantum Design) which allows rotations about the longitudinal axis and about a transverse axis. It consists of a $6\times1.5\times0.6$ mm$^3$ brass piece attached to a pulley operated by a 80 $\mu$m gold wire, see Fig.~{\ref{esquema}}(a). 
For both rotation axes, the orientation may be specified with a precision of 0.1$^\circ$, the reproducibility being $\pm0.5^\circ$ about the magnetic field axis and $\pm1^\circ$ about the transverse axis.

Before each set of measurements some operations are performed to initialize the rotating sample holder. First, $\vec m_T$ is aligned with the transverse SQUID coils. For that, the sample is zero-field cooled well below $T_c$ with its $c$ axis tilted with respect to the magnetic field direction, and a magnetic field in the Oe range (well below the lower critical magnetic field) is applied. The plate-like shape of the sample makes the demagnetizing effect highly anisotropic, giving rise to an observable $\vec m_T$. Its alignment with the transverse SQUID axis was obtained by rotation about the longitudinal axis until a maximum in the transverse SQUID was detected [see Fig.~1(b)].
The second step is to determine the correspondence between the pulley orientation and the angle between the crystal $c$ axis and the applied magnetic field, $\theta$. For that, $m_L$ is measured against the pulley orientation and compared with the expected $m_L(\theta)$ dependence in the Meissner region,\cite{yaron,mosqueira99}
\begin{equation}
m_L(\theta)\approx -VH\left(\frac{\cos^2\theta}{1-D_\perp}+\frac{\sin^2\theta}{1-D_\parallel}\right).
\label{mltheta}
\end{equation}
Here $D_\perp$ and $D_\parallel$ are the demagnetizing factors in the directions perpendicular and parallel to the CuO$_2$ layers, and $V$ is the sample volume. As $1>D_\perp\gg D_\parallel>0$, the minimum in the $m_L$ angular dependence corresponds to $\theta=0^\circ$. An example of application of this procedure is presented in Fig.~1(c). Let us finally mention that to avoid a $\sim3^\circ$ backlash in the pulley mechanism, we always set angular positions by rotating in the same direction.

%
%
\begin{figure}[t]
\includegraphics[scale=.5]{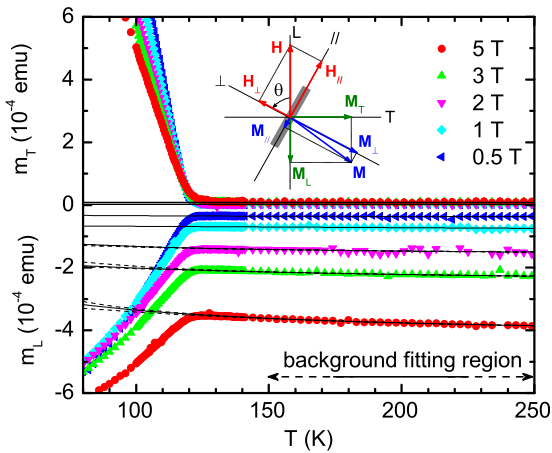}
\caption{(Color online) Determination of the sample holder and normal-state contributions to $m_L$ and $m_T$. The measurements in this example where performed with the crystal $c$ axis tilted $\theta=80^\circ$ with respect to the applied magnetic field. The lines are fits of a Curie function (in the case of $m_L$) and of a constant (in the case of $m_T$) to the data between $\sim150$ and $\sim250$ K. The dashed lines represent the uncertainty in the background determination. The diagram indicates the crystal orientation with respect to the ($\perp$,$\parallel$) and ($L$,$T$) axes (the crystal $c$-axis is parallel to the $\perp$-axis).}
\label{backs}
\end{figure}

We measured $m_L$ and $m_T$ against temperature in the reversible region of the mixed state by using different constant $\theta$ and $H$ values. To characterize with accuracy the contribution of the sample holder to the magnetic moment, the measurements were extended well above $T_c$, up to $\sim250$~K. An example corresponding to the measurements with $\theta=80^\circ$ is presented in Fig.~\ref{backs}, where the choice for the ($\perp$,$\parallel$) and ($L$,$T$) axes is also indicated. As may be seen, $m_L$ presents a diamagnetic contribution above $T_c$ mainly coming from the sample holder which is of the same order of magnitude than the one of the sample in the mixed state. It was removed by subtracting to the data a Curie-like function [$m_L=a+c/(T-b)$, where $a$, $b$ and $c$ are constants] fitted in a temperature interval well above $T_c$ (typically between 150~K and 250~K), where even the effect of superconducting fluctuations is negligible. Finally, the transverse magnetic moment above $T_c$ presents a temperature independent signal, of the order of the instrumental sensitivity, which was also subtracted.

%
%

\begin{figure}[t]
\includegraphics[scale=0.48]{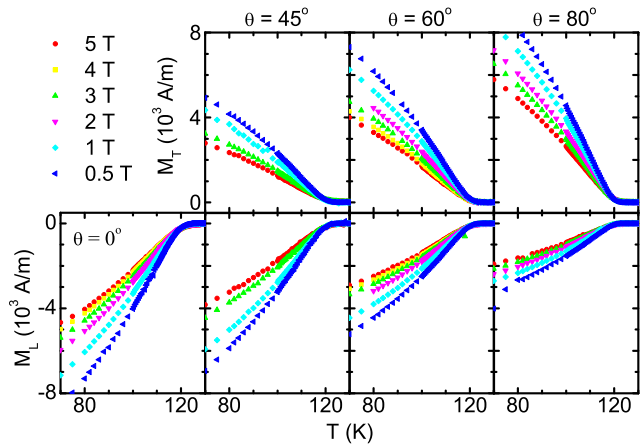}
\caption{(Color online) Temperature dependence of the longitudinal and transverse components of the magnetization (already corrected for the background contribution) for different magnetic field amplitudes and orientations.}
\label{paneles}
\end{figure}

An overview of the resulting longitudinal and transverse \textit{superconducting} magnetizations ($M_L$ and $M_T$, respectively) is presented in Fig.~\ref{paneles} as a function of temperature and for different magnetic field amplitudes and orientations. As expected in view of the high anisotropy of the compound under study, at a given temperature and magnetic field $|M_T|$ increases and $|M_L|$ decreases with $\theta$, to the point that $|M_T|$ even exceeds the corresponding $|M_L|$ value for $\theta>45^\circ$.

%
%
\begin{figure}[t]
\includegraphics[scale=.63]{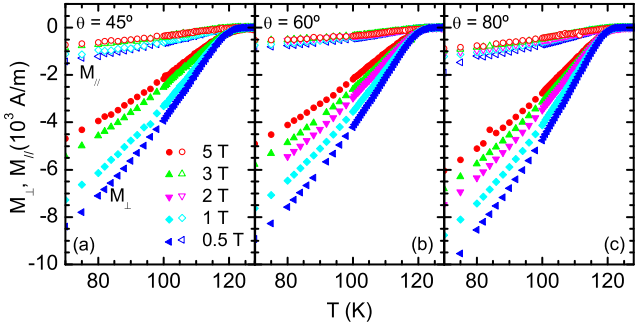}
\caption{(Color online) Temperature dependence of $M_\parallel$ (open symbols) and $M_\perp$ (closed symbols) for different magnetic field amplitudes and orientations, as results of applying Eq.~(\ref{matrix}) to the $M_L$ and $M_T$ data in Fig.~\ref{paneles}.}
\label{paneles2}
\end{figure}

\section{Data analysis}

In a recent work, we have measured the angular dependence of the magnetization vector in the same crystal for temperatures close to $T_c$ ($T/T_c\stackrel{>}{\sim}0.8$) under a 1 T magnetic field, well within the London regime, $H_{c1}^\perp\ll H_\perp\ll H_{c2}^\perp$.\cite{angular} It was shown that the $\vec M$ components parallel and perpendicular to the CuO$_2$ layers verify Eqs.~(\ref{eq1}) and (\ref{eq2}) within the experimental uncertainty, which implies that the magnetic response is largely due to the $\vec H$ component perpendicular to the CuO$_2$ layers. Such a behavior extends up to $\theta$ values 0.3$^\circ$ away from 90$^\circ$, which is consistent with a lower bound for the anisotropy factor of $\gamma\sim200$.\cite{angular} Here we check whether this may be generalized to the different regions in the reversible mixed state. For that, $M_\perp$ and $M_\parallel$ were obtained from the $M_L$ and $M_T$ data in Fig.~\ref{paneles} through 
\begin{equation}
\left(
\begin{array}{c}
M_{\perp}\\
M_\parallel
\end{array}
\right)
=
\left(
\begin{array}{cc}
\cos\theta & -\sin\theta \\
\sin\theta & \cos\theta  
\end{array}
\right)
\left(
\begin{array}{c}
M_L\\
M_T
\end{array}
\right).
\label{matrix}
\end{equation}
The result is presented in Fig.~\ref{paneles2}. As it is clearly seen already in this figure, the $M_\perp$ amplitude increases slightly with $\theta$, which could be associated with the corresponding reduction in the perpendicular component of the applied magnetic field $H_\perp=H\cos\theta$.
In what concerns $M_\parallel$, close to $T_c$ (typically above 110 K) its amplitude is within the experimental uncertainty associated with the rotating sample holder, but it grows beyond such uncertainty on lowering the temperature. In particular, for $T=70$ K$\approx0.6T_c$, when $\theta=45^\circ$ (the orientation at which $H_\perp=H_\parallel$) the $M_\parallel$ amplitude is just one order of magnitude smaller than the one of $M_\perp$. These results will be analyzed in detail in the next subsections in different regions of the reversible mixed state.

%
%

\begin{figure}[b]
\includegraphics[scale=0.35]{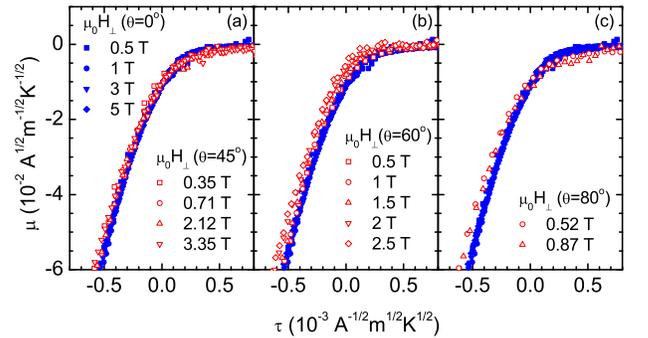}
\caption{(Color online) Scaling of the $\mu(\tau)_{H_\perp}$ curves in the critical fluctuation region close to $T_c$ as evaluated from Eqs.~(\ref{scalingm}) and (\ref{scalingt}) with $\mu_0H_{c2}(0)=340$~T (see Ref.~\onlinecite{Tl2223}). Red open symbols were obtained from measurements performed with the magnetic field tilted from the $c$ axis 45$^\circ$ (a), 60$^\circ$ (b), and 80$^\circ$ (c). For comparison, data obtained with $H\perp ab$ (i.e., $\theta=0^\circ$) are also presented (blue solid symbols).}
\label{scaling}
\end{figure}

\subsection{Analysis in the critical fluctuation region, $H_\perp\sim H_{c2}^\perp(T)$}

This region, dominated by important fluctuations of the superconducting order parameter, is bounded by the so-called field dependent Ginzburg criterion,\cite{criterion}
\begin{equation}
\frac{|T-T^\perp_c(H_\perp)|}{T_c}\approx\sqrt{\frac{4\pi k_B\mu_0}{\phi_0s\Delta c}\frac{H_\perp}{H_{c2}^\perp(0)}},
\label{criterion}
\end{equation}
where $T_c^\perp(H_\perp)=T_c[1-H_\perp/H_{c2}^\perp(0)]$, $H_{c2}^\perp(0)$ is the linear extrapolation to $T=0$~K of the perpendicular upper critical field, $\Delta c$ the specific heat jump at $T_c$, $s=1.78$~nm the CuO$_2$ layers periodicity length, $\phi_0$ the flux quantum, $k_B$ the Boltzmann constant, and $\mu_0$ the magnetic permeability. By using superconducting parameters typical of Tl-2223,\cite{Tl2223} even for the largest magnetic field amplitudes studied in this work (5~T) the critical region extends down to $\sim$110 K. So close to $T_c$ it is found $M_\parallel\approx 0$ within the experimental uncertainty.
In what concerns $M_\perp$, the GL approach for highly anisotropic superconductors predicts a scaling behavior $\mu=\mu(\tau)$, where  
\begin{equation}
\mu\equiv \frac{M_\perp}{\sqrt{H_\perp T}}
\label{scalingm}
\end{equation}
is the \textit{scaled magnetization}, and
\begin{equation}
\tau\equiv \frac{T-T_c^\perp(H_\perp)}{\sqrt{H_\perp T}}
\label{scalingt}
\end{equation}
the \textit{scaled temperature}.\cite{ullah} In a previous work we have shown the validity of this scaling in data obtained with $H\perp ab$ in the same crystal by using $\mu_{0}H_{c2}^\perp(0)=340$~T (a value resulting from the analysis of the fluctuation diamagnetism above $T_c$).\cite{Tl2223} The scaled data were shown to be also in excellent agreement with the scaling function calculated by Te\u{s}anovi\'{c} and coworkers\cite{tesanovic} when evaluated with the same $\mu_0H_{c2}^\perp(0)$ value.
Now the validity of such a scaling is checked by using the $M_\perp(T,H)$ data in Fig.~\ref{paneles2}, which were obtained with $\vec H$ tilted $\theta=45^\circ$, $60^\circ$ and $80^\circ$ from the crystal $c$ axis. The result is presented in Fig.~\ref{scaling}, where for comparison the data obtained with $\theta=0^\circ$ are also included. As may be clearly seen, the data obtained with $\theta\neq0^\circ$ are affected by a larger dispersion associated with the determination of the sample holder background. In spite of that, these data still scale according to Eqs.~(\ref{scalingm}) and (\ref{scalingt}), and are in good agreement with the scaling of the data obtained with $\theta=0^\circ$. These results confirm the validity of Eqs.~(\ref{eq1}) and (\ref{eq2}) in the critical region, which implies that near $T_c$ the magnetic response is essentially due to the $\vec H$ component perpendicular to the CuO$_2$ layers.

\subsection{Analysis in the London region of the reversible mixed state, $H_{c1}^\perp(T)\ll H_\perp\ll H_{c2}^\perp(T)$}

\begin{figure}[t]
\includegraphics[scale=.55]{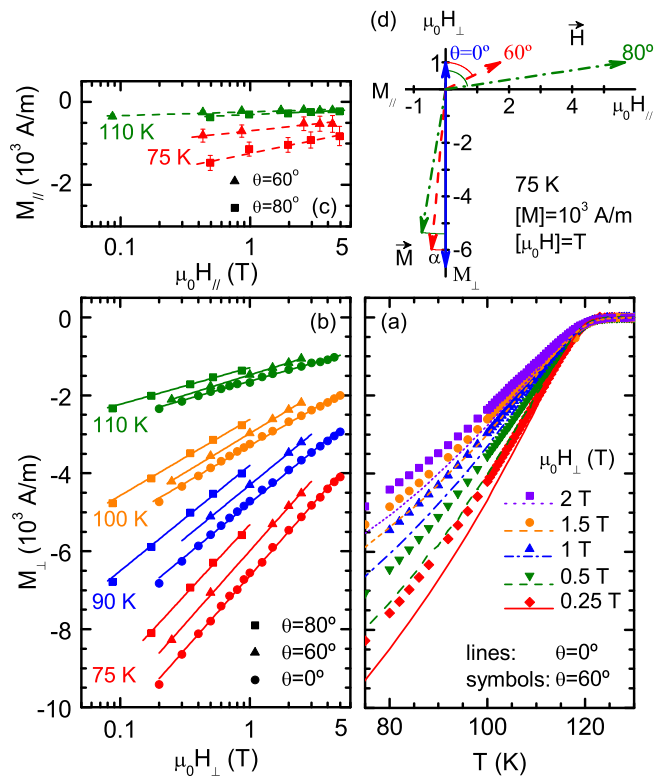}
\caption{(Color online) (a) Example of temperature dependence of $M_\perp$ below $T_c$ for some constant $H_\perp$ values. The lines were obtained with $\theta=0^\circ$ (and then with $H_\parallel=0$), while the data points were obtained with $\theta=60^\circ$ (and then with $H_\parallel=H_\perp\tan\theta\neq0$). (b) $H_\perp$ dependence of $M_\perp$ for different temperatures below $T_c$ and for different crystal orientations (and then, different $H_\parallel$ values). In this representation it is clearly shown that the presence of $H_\parallel=H_\perp\tan\theta$ reduces the $M_\perp$ amplitude by displacing vertically the curves. The lines are fits to Eq.~(\ref{blk}). In (c) it is shown that such an effect is accompanied by an increase of $|M_\parallel|$ up to observable values (for clarity only two isotherms are shown). The lines are fits to a London-like function, $M_\parallel=cte+cte'\ln H_\parallel$. (d) Example (for $T=75$ K) of the $\vec M$ deviation from the $c$ axis as $H_\parallel$ is increased keeping $H_\perp$ constant. }
\label{rotation}
\end{figure}

Almost all data points below 100~K in Fig.~\ref{paneles2} are well within this region. As commented above, here $M_\parallel$ present amplitudes clearly above the experimental uncertainty. Now we will check whether $M_\perp$ is still determined by the perpendicular component of the applied magnetic field, $H_\perp$, or presents deviations with respect to this behavior. In Fig.~\ref{rotation}(a) we present some examples of the temperature dependence of $M_\perp$ for some constant $\mu_0H_\perp$ values in the range 0.25-2 T. These data come from measurements obtained with $\theta=60^\circ$, so that the parallel component of the applied field is $H_\parallel=H_\perp\tan60^\circ\approx1.72H_\perp$. For comparison, we include measurements obtained with the same magnetic field values but applied directly perpendicular to the \textit{ab} layers so that $H_\parallel=0$. As may be clearly seen, $|M_\perp(T,H_\perp,H_\parallel)|$ is systematically smaller than $|M_\perp(T,H_\perp,0)|$, the difference increasing on lowering the temperature below $T_c$ up to values well above the experimental uncertainty. For instance, at $T=80$~K, even for the largest $\mu_0H_\perp$ values (2 T), this uncertainty is about 5\% (close to the data points size), while the difference between $|M_\perp(T,H_\perp,H_\parallel)|$ and $|M_\perp(T,H_\perp,0)|$ is above 20\%.

%
%

To go deeper into this effect, in Fig.~\ref{rotation}(b) we present the $H_\perp$ dependence of $M_\perp$ as determined from measurements obtained under different crystal orientations (and then, under different $H_\parallel$ values). For comparison, we have also included measurements performed with $\theta=0^\circ$ (and then, with $H_\parallel=0$). As predicted by London-like approaches,\cite{tinkham} $M_\perp$ presents a linear behaviour with respect to $\ln H_\perp$. As may be clearly seen, in agreement with Fig.~\ref{rotation}(a) the $M_\perp(H_\perp)$ curves taken at the same temperature are displaced to lower amplitudes as $\theta$ (or $H_\parallel$) is increased, but without changing appreciably the slope. 

In Fig.~\ref{rotation}(c) we present the corresponding $M_\parallel$ dependence on $H_\parallel$. In spite of the experimental uncertainties affecting these data (mainly associated with the rotating sample holder), $|M_\parallel|$ seems to increase up to observable values on increasing $\theta$. As shown in Fig.~\ref{rotation}(d) this represents a slippage of $\vec M$ from the crystal $c$ axis. At a temperature of 75 K, the angle between $\vec M$ and the $c$ axis, $\alpha$, is as high as $\sim8^\circ$ when $\theta=80^\circ$, which is orders of magnitude larger than the one predicted by the conventional anisotropic GL approach: $\alpha=\arctan(\gamma^{-2}\tan\theta)$, where $\gamma$ is the anisotropy factor. In fact, taking for this material $\gamma\stackrel{>}{_\sim}200$ (see e.g., Ref.~\onlinecite{angular}) one finds $\alpha\stackrel{<}{_\sim}0.01^\circ$ for $\theta=80^\circ$. 

\begin{figure}[b]
\includegraphics[scale=.5]{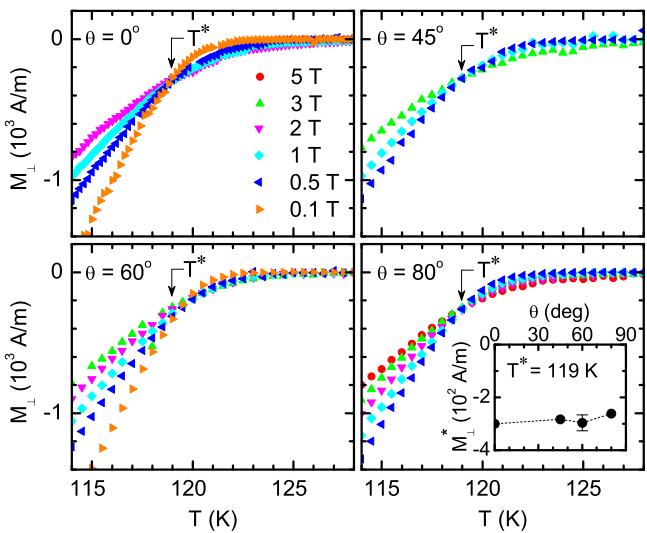}
\caption{(Color online) Detail around $T_c$ of the $M_\perp(T)_H$ curves obtained for different crystal orientations relative to the applied field. The crossing point at a temperature few degrees below $T_c$ is clearly seen. In the inset it is shown that $M^*_\perp$ is almost independent of $\theta$.}
\label{fig_crossing}
\end{figure}

Now we show that the $M_\perp(H_\perp)$ dependence on $\theta$ may be phenomenologically explained in the framework of the well known approach by Bulaevskii, Ledvig and Kogan (BLK) for the perpendicular magnetization of highly anisotropic layered superconductors in the London regime:\cite{BLK} 
\begin{eqnarray}
M_\perp=-M_0\ln\left(\frac{\eta H_{c2}^\perp}{H_\perp}\right)
+M_1\ln\left(\frac{M_1}{M_0C}\frac{\eta H_{c2}^\perp}{H_\perp}\right),
\label{blk}
\end{eqnarray}
Here $M_0=f\phi_0/8\pi\mu_0\lambda_{ab}^2$, $M_1=fk_BT/\phi_0s$, $C$ and $\eta$ constants of the order of the unity, $\lambda_{ab}$ the magnetic penetration length in the $ab$ planes, and $f$ the effective superconducting fraction (which may be approximated by the Meissner fraction).\cite{volume_fraction} The first term on the right is the conventional London magnetization, whereas the second one is associated with thermal fluctuations of the two-dimensional vortex (\textit{pancakes}) positions. A direct consequence of Eq.~(\ref{blk}) is the crossing of the $M_\perp(T)_{H_\perp}$ curves at a temperature $T^*$ a few degrees below $T_c$, the crossing point magnetization being
\begin{equation}
M_\perp^*=-f\frac{k_BT^*}{\phi_0s}\ln C.
\label{eq_crossing}
\end{equation}
A first comparison with the experimental data may be done through the crossing point coordinates. In Fig.~\ref{fig_crossing} we present a detail around $T_c$ of $M_\perp(T)_H$ measured with different $\theta$ values, where the crossing point is clearly seen. The $H_\perp$ values in this figure are well below $H_{c2}^\perp(T^*)$, so that Eq.~(\ref{eq_crossing}) is applicable. The analysis shows that $M^*_\perp$ is almost $\theta$ independent up to $\theta=80^\circ$, the scattering in the $M_\perp^*$ values (about 10\%) being well within the uncertainty associated with the background at these temperatures. The comparison with Eq.~(\ref{eq_crossing}) allowed to determine $C\approx2.2$.

%
%

%
%
\begin{figure}[t]
\includegraphics[scale=0.5]{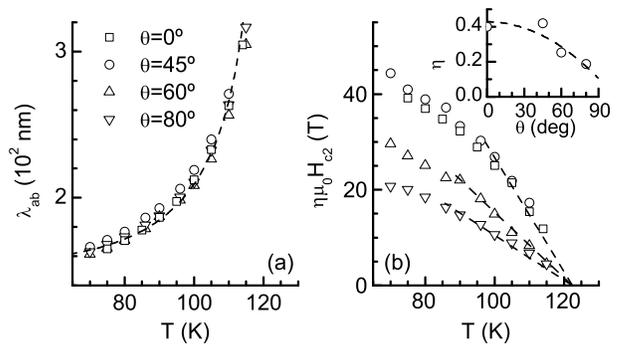}
\caption{Temperature dependences of $\lambda_{ab}$ (a) and of $\eta H_{c2}$ (b) as follows from the comparison of Eq.~(9) with the $M_\perp(H_\perp)_\theta$ data. By using the $H_{c2}(T\to0)$ value obtained in Ref.~\onlinecite{Tl2223}, it follows the $\eta(\theta)$ dependence shown in the inset. The lines in (a) and (b) are fits to $\lambda_{ab}(T)=\lambda_{ab}(0)/\sqrt{1-(T/T_c)^n}$ and $H_{c2}(T)=H_{c2}(0)(1-T/T_c)$ near $T_c$, respectively.}
\label{parametros}
\end{figure}

A thorough comparison of Eq.~(\ref{blk}) with the experimental data is presented in Fig.~\ref{rotation}(b). The solid lines are fits of Eq.~(\ref{blk}) to each isotherm, with $C=2.2$ and $\eta H_{c2}$ and $\lambda_{ab}$ as free parameters. The resulting $\eta H_{c2}^\perp(T)$ and $\lambda_{ab}(T)$ are presented in Fig.~\ref{parametros}. The $\lambda_{ab}(T)$ curves coming from data under different $\theta$ values agree with each other, which is an important check of the applicability of Eq.~(\ref{blk}). In turn, $\eta H_{c2}^\perp$ follows a linear temperature dependence close to $T_c$, in full agreement with the conventional (GL) behavior.\cite{notakogan} The differences between the $\eta H_{c2}^\perp(T)$ coming from measurements under different $\theta$ values may be interpreted by assuming that $\eta$ presents a slight dependence on $\theta$. 
As commented above, the analysis of the fluctuation magnetization around $T_c$ with $H\perp ab$ in this crystal leads $\mu_0H_{c2}^\perp(0)=340$ T.\cite{Tl2223} By combining this value with the data in Fig.~\ref{parametros}(b), we obtained the angular dependence of $\eta$ shown in the inset of that figure.

A theoretical approach for the angular dependence of $M_\perp$ and $M_\parallel$ in the framework of the Lawrence-Doniach (LD) theory including the effect of thermal fluctuations and of the multilaminarity is not available. However, the effects associated with thermal fluctuations decrease rapidly on lowering the temperature below $T_c$. As an example, at the lower temperature studied in this work (75~K) the thermal fluctuations contribution to $M_\perp$ in Eq.~(\ref{blk}) is less than 5\% the of the total $M_\perp$ amplitude. Then, the LD model for single layered HTSC without corrections associated with thermal fluctuations should be a good approximation well below $T_c$ provided that the three closest layers in the periodicity length behave as an unique superconducting layer. This model was used by Feinberg\cite{feinberg} to calculate $M_{\perp}(T,H)$ and $M_{\parallel}(T,H)$ as a function of the orientation of the applied magnetic field in the London region. In that work it is shown that in the case of highly anisotropic superconductors, three angular regimes may be distinguished:\cite{bulaevskii92} 

1) $\tan\theta<\xi_{ab}/s$ ($\xi_{ab}$ is the \textit{in-plane} superconducting coherence length). In this region, the separation (in the direction of the layers) between 2D vortices in adjacent layers $\ell=s\tan\theta$ is smaller than their core size (i.e., $\ell<\xi_{ab}$). The effective vortex cores are identical to the ones of the 3D anisotropic case, and a 3D anisotropic description of the mixed state is directly applicable.

2) $\xi_{ab}/s<\tan\theta<\gamma$. In this region the distance between 2D vortices in adjacent layers is beyond the cores size (i.e., $\ell>\xi_{ab}$), although smaller than the {\it Josephson bending length} $r_j=\gamma s$. The 3D anisotropic approach is still valid, but with a modified cutoff to account for the effective core dimensions (of the order of $\ell$). The effect on $M_\perp$ may be implemented by just including a constant in the logarithmic dependence of $H_\perp$. 

3) $\tan\theta>\gamma$. Here $\ell>r_j$ and Josephson vortices develop between 2D vortices in adjacent layers. As the anisotropy factor in the sample studied is larger than $\gamma=200$, this region is restricted to $\theta$ values $\sim0.3^\circ$ (or less) away from 90$^\circ$. 

According to the scenario summarized above, the observed angular dependence of the parameter $\eta$ above $\theta=45^{\circ}$ could be attributed to a transition from region 1 to region 2. In fact, by using $\xi_{ab}(T\to0)=1.0$ nm,\cite{Tl2223} the $\theta$ value separating both regions is $\theta_{1\to2}=\arctan(\xi_{ab}/s)\approx42^\circ$ at 75 K, a value compatible with the onset of the above mentioned anomalies.
In what concerns the results for $M_{\parallel}(T,\vec H)$, the LD model for single-layered compounds predicts a contribution of the order of $M_{\perp}\tan\theta/\gamma^2$, which is orders of magnitude smaller than observed.\cite{feinberg} We then suggest that the disagreement could be a consequence of the multilayered nature of the compound studied (for instance, associated with a stronger Josephson coupling between the three closest layers in the periodicity length).

%
%

\begin{figure}[b]
\includegraphics[scale=0.5]{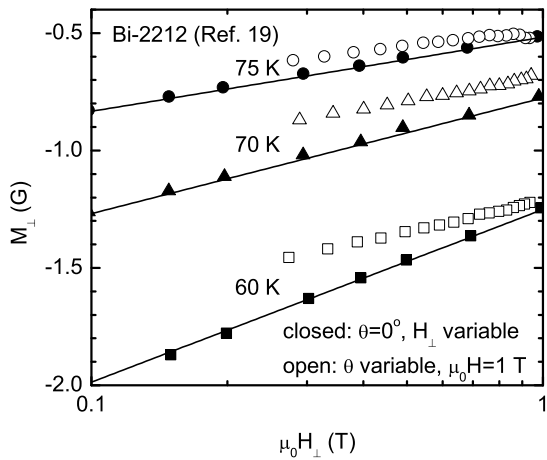}
\caption{$H_\perp$ dependence of $M_\perp$ in the mixed state of Bi-2212, as follows from data in Ref.~\onlinecite{tuominen1}. Solid data points were obtained directly with $H\perp ab$, while open data points were derived [by using Eq.~(\ref{matrix})] from the angular dependence of $M_L$ and $M_T$ in the presence of a 1 T magnetic field. The lines are fits to Eq.~(\ref{blk}).}
\label{tuominen}
\end{figure}

\subsection{Possible confirmations of the observed anomaly in other multilayered HTSC in the literature}

By inspecting the literature some examples may be found which confirm the breakdown of Eqs.~(1) and (2) in the mixed state of other highly anisotropic multilayered HTSC. The first corresponds to the pioneering work by Tuominen \textit{et al.}\cite{tuominen1} where simultaneous measurements of $M_L$ and $M_T$ were presented for the first time in a HTSC (Bi$_2$Sr$_2$CaCu$_2$O$_8$, hereafter Bi-2212). In Fig.~\ref{tuominen} we present $M_\perp(H_\perp)$ data for different temperatures below $T_c$ coming from this work. Closed symbols were measured directly with $\theta=0^\circ$, while open symbols were obtained from measurements of $M_L(\theta)$ and $M_T(\theta)$ in the presence of a constant applied magnetic field (1 T). As may be clearly seen, the $M_\perp(H_\perp)$ amplitude is progressively reduced with $\theta$, the effect being more pronounced in the isotherms corresponding to lower temperatures. This effect is in good qualitative agreement with our present results. It is worth noting, however, that the anisotropy factor found for the Bi-2212 sample in this work is anomalously small ($\gamma\approx17$) which may indicate the presence in the sample of structural inhomogeneities.

%
%

\begin{figure}[t]
\includegraphics[scale=0.5]{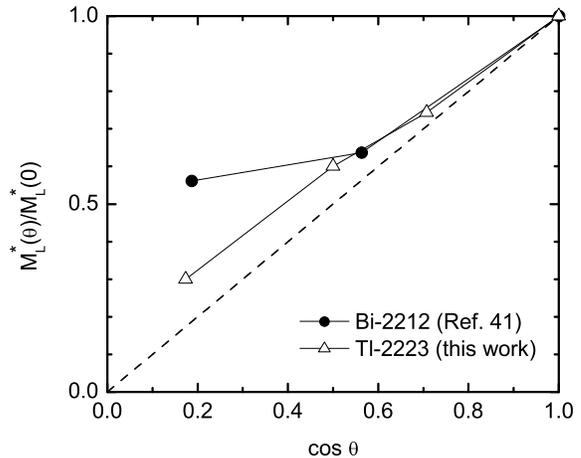}
\caption{Angular dependence of the longitudinal magnetization at the crossing point temperature. The dashed line corresponds to the dependence $M_L^*\propto\cos\theta$ based on the applicability of Eqs.~(\ref{eq1}) and (\ref{eq2}). }
\label{cruceBi2212}
\end{figure}

Measurements of the angular dependence of $M_L$ alone may also be used to confirm the applicability of Eqs.~(\ref{eq1}) and (\ref{eq2}). In particular, if they are valid it may be approximated
\begin{equation}
M_L(\theta,H)\approx M_\perp(H_\perp)\cos\theta.
\end{equation}
At the crossing point temperature $M_\perp$ is independent of $H_\perp$, and $M_L$ should be then proportional to $\cos\theta$. This dependence was probed by Li \textit{et al.}\cite{qiangli} in a Bi-2212 single crystal, obtaining the notable deviation shown in Fig.~\ref{cruceBi2212} (closed symbols). Such a behavior was left unexplained, but it is qualitatively similar to the one of our Tl-2223 crystal (open symbols), which is due to the non-negligible contribution associated with the parallel component of the magnetization vector, $M_\parallel\sin\theta$.

\begin{figure}[t]
\includegraphics[scale=0.5]{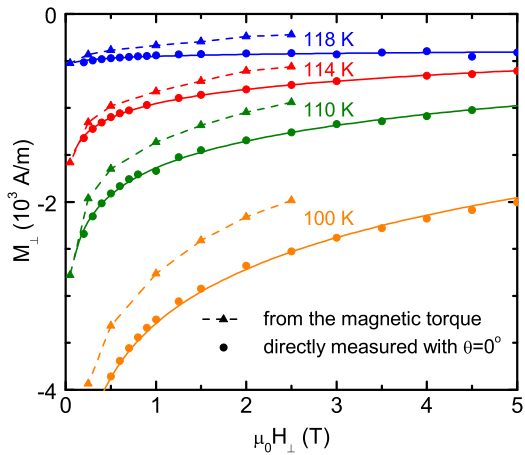}
\caption{Comparison between the magnetic field dependences of $M_\perp$ as measured directly with $H\perp ab$ (circles) and as results from the magnetic torque by using Eq.~(\ref{aproxtorque}) (triangles). In turn, torque data were obtained from the $M_T$ data in Fig.~\ref{paneles} corresponding to $\theta=60^\circ$. Solid lines are fits to Eq.~(\ref{blk}), and dashed lines guides for the eyes. }
\label{torque}
\end{figure}

An additional proof of the breakdown of Eqs.~(\ref{eq1}) and (\ref{eq2}) in Bi-2212 crystals may be found in the high-field measurements of the magnetic torque by Li \textit{et al.}\cite{li05} By assuming the applicability of these equations, these authors obtained the perpendicular component of the magnetization vector from the magnetic torque $\tau$ through
\begin{equation}
M_\perp=\frac{\tau}{\mu_0 H\sin\theta}.
\label{aproxtorque}
\end{equation}
In Fig.~3 in their paper, so obtained $M_\perp(H_\perp)$ isotherms are compared with SQUID measurements performed directly with $H\perp ab$. While for temperatures about and above $T_c=86$~K both measurements overlap, below 85~K the $M_\perp$ data coming from torque magnetometry (obtained with the crystal $c$-axis tilted 15$^\circ$ with respect to the applied field) present a slightly \textit{smaller} amplitude. This result is consistent with our present findings, as may be clearly seen in Fig.~\ref{torque} where the same comparison is done with our data for Tl-2223. In this case $\tau$ was obtained from the transverse component of the magnetization vector (presented in Fig.~\ref{paneles}) through $\tau=-\mu_0HM_T$. It is worth noting that the difference between the $M_\perp(H_\perp)$ data resulting from torque and SQUID measurements differ not only in the amplitude, but also in the $H_\perp$ dependence (a fact which cannot be appreciated in Ref.~\onlinecite{li05} because the SQUID measurements extend only up to $\mu_0H_\perp=0.1$~T). In particular, the high-field crossing point observed at 118 K (note the $H_\perp$ independence of $M_\perp$ at this temperature), is not observed in the $M_\perp(H_\perp)$ data coming from the magnetic torque.

%
%

\subsection{Implications on the $M_\perp(T,H_\perp)$ derived from the magnetic torque in highly anisotropic multilayered superconductors}

Torque magnetometry allows to investigate with a high resolution the magnetic response of anisotropic samples in the presence of large applied magnetic fields. In some recent works this technique was used to study the magnetization of several HTSC families under fields up to 20~-~45~T.\cite{li05,li10} In these works, on the basis of the large anisotropy of the compounds studied, Eq.~(\ref{aproxtorque}) was used to approximate $M_\perp$ from the magnetic torque. The resulting temperature and magnetic field dependences of $M_\perp$ were claimed to be beyond conventional Ginzburg-Landau descriptions for the effect of thermal fluctuations. 
In particular, below $T_c$ the $M_\perp(H_\perp)$ isotherms present a pervasive nonlinearity, to the point that the crossing point typical of highly anisotropic HTSC disappears in the presence of magnetic fields larger than $\sim5$ T.  
On the other side, $M_\perp$ presents a strongly non-linear diamagnetic response up to temperatures well above $T_c$.

In what concerns the behavior below $T_c$, the results summarized in our present work pose serious doubts on the applicability of the approximations leading to Eq.~(\ref{aproxtorque}). As a consequence, the perpendicular magnetization deduced from torque measurements may not be representative of the intrinsic $M_\perp(H_\perp)$ behavior. 
Further measurements would be needed to decide whether the disappearance of the crossing point is intrinsic or an artifact due to a non negligible contribution to the magnetic torque coming from $M_\parallel$. For instance, measurements of the magnetic field dependence of the magnetic torque under different crystal orientations would allow to probe the applicability of Eqs.~(1) and (2). 

Regarding the behavior above $T_c$, the overlapping of the $M_\perp(H_\perp)$ data coming from SQUID and torque magnetometry suggests that Eq.~(\ref{aproxtorque}) may be applicable in this region. However, a detailed analysis of data for optimally-doped Bi-2212 in this region,\cite{comment} showed that the observed non-linear diamagnetic response above $T_c$ follows closely the prediction of the Gaussian GL approach under a \textit{total energy cutoff}, including the vanishing of fluctuation effects at $T\approx 1.7T_c$. The same applies also to underdoped Bi-2212 if the effect of recognized $T_c$ inhomogeneities (which in these materials could even have an \textit{intrinsic} origin)\cite{intrinsicinh} is taken into account.\cite{comment} Recent measurements in high quality samples of several HTSC families fully confirm these conclusions.\cite{nuestros,nuestroY123,nuestroBi2212}

\section{Conclusions}

We have presented detailed measurements of the magnetization vector $\vec M$ in the mixed state of a high quality Tl-2223 crystal in the presence of tilted magnetic fields. In the critical region close to $T_c$ the response expected for highly anisotropic HTSC is observed within the experimental uncertainty:  $\vec M$ is perpendicular to the CuO$_2$ (\textit{ab}) layers, and its amplitude is only dependent on the component of the applied magnetic field in the same direction ($H_\perp$).
However, farther below $T_c$, in the London region, a clear deviation from this behavior is observed: the perpendicular component of $\vec M$ ($M_\perp$) shrinks on tilting the magnetic field from the crystal $c$ axis while leaving $H_\perp$ constant. In turn, the parallel component ($M_\parallel$) grows above the experimental uncertainty. 
The $M_\perp$ behavior is phenomenologically explained in terms of the well known BLK approach for highly anisotropic HTSC, by just assuming that the so-called vortex structure constant, $\eta$, is slightly dependent on the magnetic field orientation. Such a dependence is justified by the corresponding dependence on the field orientation of the effective size of the vortex cores. Regarding $M_\parallel$, its amplitude is orders of magnitude larger than predicted by theoretical approaches for single layered HTSC, which led us to suggest that it could be a consequence of the multilayered structure of the compound under study. 

The analysis of magnetization and torque data in the literature shows that the same effect could also be present in the mixed state of other highly anisotropic HTSC (Bi-2212). If confirmed, such an effect may have implications in the interpretation of recent measurements of the magnetic torque in highly anisotropic HTSC in terms of $M_\perp$.\cite{li10} In particular, features like the vanishing of the crossing point under large magnetic field amplitudes,\cite{naughton} could be an artifact associated with not taking into account the $M_\parallel$ contribution to the magnetic torque.

It would be interesting to extend the present measurements to other HTSC with different number of CuO$_2$ layers in the periodicity length, and also to extend to multilayered superconductors the existing approaches for the magnetization vector in the presence of tilted magnetic fields.

\section*{Acknowledgments}

This work was supported by the Spanish MICINN and ERDF \mbox{(FIS2010-19807)}, and the Xunta de Galicia (2010/XA043 and 10TMT206012PR).  We acknowledge J. Ponte his valuable help with the rotating sample holder.

\end{document}